# Energy-efficient spin Hall nano-oscillators using near-compensated CoGd ferrimagnets

*Jiayu Lei*[1], *Raghav Sharma*[2], *Shishun Zhao*[1], *Fanrui Hu*[1], *Yuchen Pu*[1], *Chenhui Zhang*[1], *Rahul Mishra*[3], and *Hyunsoo Yang*[1] *

[1]Department of Electrical and Computer Engineering, National University of Singapore, Singapore 117583, Singapore

[2]Department of Electrical Engineering, Indian Institute of Technology Ropar, Rupnagar, India

[3]Center for Applied Research in Electronics, Indian Institute of Technology Delhi, Delhi 110016, India

*Corresponding author: eleyang@nus.edu.sg

J. Lei, R. Sharma, and S. Zhao contributed equally to this work.

**Conventional spin Hall nano-oscillators (SHNOs) based on ferromagnets face practical limitations due to high threshold current densities and large external magnetic field requirements. Ferrimagnets provide an attractive alternative due to their unique magnetic dynamics and potential for energy-efficient spintronic devices. In this study, we report rare-earth−transition-metal (RE−TM) ferrimagnetic SHNOs utilizing $Co_{1-x}Gd_x$ alloys, in which compositional tuning enables high-performance operation near the magnetization compensation. The optimized SHNO operates at a low current density ($1.43 \times 10^7$ A/cm$^2$), a small magnetic field (5 mT), and exhibits a narrow linewidth (0.61 MHz) simultaneously, showing an order-of-magnitude improvement over its ferromagnetic counterparts. This enhanced performance arises from high spin-orbit torque efficiency, low magnetic anisotropy, reduced effective magnetization, and minimized nonlinearity near the compensation point. These results establish RE−TM ferrimagnets as a promising material platform for next-generation spintronic devices and offer new strategies for realizing energy-efficient, high-performance spintronic oscillators.**

## 1. Introduction

Spintronic devices are promising candidates for high-frequency technologies [1-5]. Among them, spin Hall nano-oscillators (SHNOs) have attracted growing interest due to potential applications in microwave generation [2, 6-8] and unconventional computing [9-10]. A typical SHNO consists of a heavy-metal spin source layer adjacent to a ferromagnetic layer [2, 7]. When an electric current passes through the heavy metal, the spin Hall effect generates a transverse spin current that exerts a spin-orbit torque (SOT) on the ferromagnet. This torque compensates damping and drives auto-oscillations of the magnetization. The resulting magnetization precession modulates the device resistance through the anisotropic magnetoresistance, producing a detectable microwave voltage signal. SHNOs stand out for their simple structures [7], nanoscale dimensions [11-12], frequency tunability [13], and compatibility with standard semiconductor fabrication technologies [14].

Despite these advantages, the practical implementation of ferromagnet-based SHNOs is challenged by several intrinsic limitations. First, they require large threshold current densities to initiate auto-oscillations, leading to substantial power consumption [15]. Second, their operation typically depends on relatively strong external magnetic fields, ranging from tens to thousands of mT [7, 16-19], which complicates on-chip integration. Third, they exhibit broad spectral linewidths originating from nonlinear phase noise [20], limiting signal coherence and stability. Various strategies have been pursued to mitigate these challenges, including optimizing device geometries [7], adjusting lateral dimensions [11-12, 21], engineering spin source layers [15-16], and using insulating substrates [12], to reduce the threshold current density. Furthermore, the development of cryogenic easy-plane anisotropy SHNOs has enabled operation under magnetic fields below tens of mT [22]. In addition, techniques such as mutual synchronization [10, 23] and nonlinearity reduction [24] have narrowed the linewidth into the MHz range. Although these strategies have led to incremental progress compared with early SHNOs [2], the current densities and magnetic fields required for practical implementation remain high. Overcoming these limitations calls for new magnetic materials and novel application strategies that exploit the unique properties of SHNOs beyond their traditional roles.

Ferrimagnets composed of two exchange-coupled rare-earth (RE) and transition-metal (TM) sublattices offer a viable route to overcome these constraints. Their magnetic properties can be precisely tuned through composition control [25-27], showing great potential in memory [25, 28] and ultrafast electronics [26, 29] due to their low damping [30], efficient SOT [25, 31-32] and spin-transfer torque [33-34], and broad resonance frequencies ranging from GHz to sub-THz exchange modes [29, 35]. Near the magnetization compensation point, RE−TM alloys such as $Co_{1-x}Gd_x$ combine small magnetization and magnetic anisotropy with strong exchange coupling, allowing efficient operation at low current densities under small magnetic fields. These characteristics position RE−TM ferrimagnets as a promising material platform for next-generation SHNOs.

In this study, we experimentally demonstrate the first near-compensated ferrimagnetic SHNOs. The optimized Pt/$Co_{73.1}Gd_{26.9}$ SHNO displays enhanced SOT efficiency, reduced anisotropy, and suppressed nonlinearity. These characteristics enable the Pt/$Co_{73.1}Gd_{26.9}$ SHNO to achieve a 16-fold lower threshold current density, a 10-fold smaller operating magnetic field, and a 16-fold narrower minimum linewidth than conventional Pt/Py ferromagnetic SHNOs. These findings highlight the potential of RE−TM alloys for realizing energy-efficient, high-performance spintronic devices.

## 2. Results and Discussion

In $Co_{1-x}Gd_x$ ferrimagnetic alloys, Co and Gd sublattices are antiferromagnetically coupled, as illustrated in Figure 1a. Accordingly, variations in their relative concentrations modulate the net magnetization and other magnetic properties, determining whether the alloy is Co- or Gd-dominant. The alloy compositions are controlled by adjusting the deposition power of Gd during co-sputtering (Methods), and characterized using scanning electron microscopy with energy-dispersive X-ray spectrometry (SEM-EDS), as detailed in Section S1, Supporting Information.

The Ta (1 nm)/Pt (6 nm)/$Co_{1-x}Gd_x$ (7 nm)/$TaO_x$ (2 nm) stacks are deposited on highly resistive Si substrates to minimize microwave losses [12], with Pt acting as the spin current source (Methods). The saturation magnetization ($M_s$) of the $Co_{1-x}Gd_x$ alloy is defined by $M_s = |M_{Co} - M_{Gd}|$, where $M_{Co}$ and $M_{Gd}$ represent the magnetization of the Co and Gd sublattices, respectively [26]. Near the magnetization compensation point ($x_{MC}$), where $M_s$ approaches zero, strong exchange coupling enhances SOT efficiency [25]. Thus, identifying $x_{MC}$ is essential for optimizing SHNO performance. To determine $x_{MC}$, we measure out-of-plane hysteresis loops at room temperature using the magneto-optical Kerr effect (MOKE) by sweeping an external magnetic field $H_z$ perpendicular to the films. As shown in Figure 1b, all compositions exhibit perpendicular magnetic anisotropy (PMA), and a reversal in loop polarity occurs across $x_{MC}$, between Gd concentrations of 24.5% and 26.9%, marking the transition from Co-rich to Gd-rich regimes. The coercive field $H_c$ diverges and reaches its maximum near $x_{MC}$, around $x \approx$ 25% - 26%, as shown in the upper panel of Figure 1c. The lower panel of Figure 1c shows saturation magnetization $M_s$ and uniaxial anisotropy energy density $K_u$ measured by vibrating sample magnetometer (VSM), consistent with prior studies on RE−TM films [36]. Both parameters reach minima near the same composition, confirming the compensation point at approximately $x \approx$ 25% - 26%. The details of VSM measurements are provided in Section S2, Supporting Information.

To investigate the current distribution in Pt/$Co_{1-x}Gd_x$ bilayers, we characterize the electrical resistivity. $Co_{1-x}Gd_x$ exhibits a resistivity of 159.0 to 357.2 μΩ·cm (Section S3, Supporting Information), about an order of magnitude higher than that of Pt (22.7 μΩ·cm). Using the measured thin-film resistivities, the parallel-resistor and Fuchs-Sondheimer models give Pt layer current fractions of 87.5% to 94.0% and 81.9% to 90.9%, respectively, indicating that most current flows through Pt with minimized current shunting through the magnetic layer, as detailed in Section S3, Supporting Information.

To evaluate the magnetization dynamics of ferrimagnetic SHNOs, the Pt (6 nm)/$Co_{1-x}Gd_x$ (7 nm) films are patterned into 150-nm-wide nano-constriction SHNOs as illustrated in Figure 2a. A direct current ($I_{DC}$) is applied along the $x$-axis while an external field ($H_{ext}$) is applied at polar angle $\theta$ and azimuthal angle $\varphi$ (Methods and Section S4, Supporting Information). All measurements are performed at room temperature. Figures 2b and 2c show the fabricated device integrated with a ground-signal-ground coplanar waveguide and an SEM image of the constriction, respectively. Conventional Pt (6 nm)/Py (5 nm)/$SiO_2$ (5 nm) SHNOs are used as ferromagnetic references. The upper panel of Figure 2d presents the auto-oscillation emission spectrum of the near-compensated Pt/$Co_{73.1}Gd_{26.9}$ SHNO with a minimal linewidth at $I_{DC}$ = 0.17 mA and $\mu_0 H_{ext}$ = 5 mT ($\theta$ = 80°, $\varphi$ = 45°). This applied current of 0.17 mA corresponds to a current density of $1.43 \times 10^7$ A/cm$^2$ (Section S5, Supporting Information).

The linewidth, fitted by a Lorentzian function, is 0.61 MHz, which is the narrowest linewidth for a single-constriction SHNO reported so far [23-24]. For comparison, the conventional Pt/Py SHNO shows a minimal linewidth of 9.63 MHz (lower panel in Figure 2d). Current-dependent measurements (Section S6, Supporting Information) further confirm that the linewidth of the near-compensated Pt/$Co_{73.1}Gd_{26.9}$ SHNO is more than an order of magnitude narrower than that of the conventional Pt/Py SHNO. As the Pt/$Co_{73.1}Gd_{26.9}$ SHNO is operated under a smaller applied magnetic field (5 mT), its auto-oscillation frequency is correspondingly reduced relative to the Pt/Py SHNO. However, the Pt/$Co_{73.1}Gd_{26.9}$ SHNO exhibits a markedly narrower linewidth, yielding a higher quality factor ($Q = f/\Delta f$). As a result, its quality factor $Q$ reaches 832 at 5 mT (Figure 2d), representing a 2.3-fold enhancement over the Pt/Py SHNO ($Q$ = 357). Across the applied field range of 50 to 100 mT, the $Q$-factor remains between 600 and 800 at higher frequency bands (Section S6, Supporting Information).

To understand the origin of this narrow linewidth in the near-compensated SHNO, we investigate the nonlinear coefficient $N$ of the system. This coefficient reveals the sign and magnitude of the nonlinear frequency shift [37], a key factor in linewidth control [24]. The nonlinear coefficient $N$ is determined by the magnitudes of the external field $|H_{ext}|$ and the effective magnetization $M_{eff}$, and is expressed as follows [38]

$$N = \frac{\omega_H \omega_M}{\omega_0}\left(\frac{3{\omega_H}^2 \cos^2\theta_{int}}{{\omega_0}^2} - 1\right), \tag{1}$$

$$\omega_0 = \sqrt{\omega_H\left(\omega_H + \omega_M \sin^2\theta_{int}\right)}, \tag{2}$$

where $\omega_H = \gamma\mu_0|H_{int}|$, $\omega_M = \gamma\mu_0|M_{eff}|$, $\gamma / 2\pi = 29.0$ GHz/T is the gyromagnetic ratio, and $\mu_0$ is the vacuum permeability. Here, $|H_{int}|$ and $\theta_{int}$ represent the magnitude and polar angle of the internal field, respectively. Angles are defined relative to the coordinate system shown in Figure 2a. When the external field is applied in-plane, the internal field simplifies to $H_{int} = H_{ext}$ and $\theta_{int} = \theta = 90°$. Under this condition, as detailed in Section S7, Supporting Information, Equation (1) can be simplified to

$$N = -\frac{\mu_0\gamma|H_{ext}|M_{eff}}{\sqrt{|H_{ext}|\left(|H_{ext}|+M_{eff}\right)}}. \tag{3}$$

Therefore, the expression of $N^2$ is

$$N^2 = \frac{\left(\mu_0\gamma|H_{ext}|M_{eff}\right)^2}{|H_{ext}|\left(|H_{ext}|+M_{eff}\right)}. \tag{4}$$

Meanwhile, the linewidth of spin-oscillators can be described by [20, 24, 37-38]

$$\Delta f = \Delta f_{thermal}\left(1 + N^2/\Gamma_{eff}^2\right), \tag{5}$$

where $\Delta f_{thermal}$, $N$, and $\Gamma_{eff}$ represent the thermal generation linewidth, the nonlinear coefficient, and the effective damping, respectively. Thus, Equation (5) shows that the linewidth broadening is governed by $N^2$, which, according to Equation (4), depends on both the magnitude of the in-plane external field and the effective magnetization, thereby determining the nonlinear contribution to the linewidth. Based on Equation (4), Figure 3 shows how the squared nonlinear coefficient $N^2$ changes with both the magnitudes of effective magnetization $|\mu_0M_{eff}|$ and the in-plane external field $|\mu_0H_{ext}|$. Since the Pt/$Co_{1-x}Gd_x$ films exhibit PMA, $M_{eff}$ is negative in this system, and its magnitude $|\mu_0M_{eff}|$ is used in the following discussion for clarity.

As shown in Figure 3a, increasing the magnitude of both effective magnetization and external field gives rise to greater $N^2$. For a clearer comparison, Figure 3b presents the variation of $N^2$ with $|\mu_0M_{eff}|$ at fixed external field values of 5, 50, 100, and 150 mT. In all cases, $N^2$ exhibits a monotonic increase with $|\mu_0M_{eff}|$. Since $\Delta f$ is proportional to $N^2$ as indicated by Equation (5), and given that $N^2$ increases almost linearly with $|\mu_0M_{eff}|$ (Figure 3b), it is expected that $\Delta f$ and $|\mu_0M_{eff}|$ share a similar trend as a function of Gd concentration $x$, as shown in Figure 3c. As $|\mu_0M_{eff}|$ and $N^2$ increase away from the

compensation point, $\Delta f$ correspondingly broadens in both Co-rich and Gd-rich SHNOs. To evaluate the relationship between the measured $\Delta f$ and calculated $N^2$, $\Delta f$ and $N^2$ as functions of Gd concentration are plotted in Figure 3d. The lowest linewidth, $\Delta f$, is observed in the Pt/$Co_{73.1}Gd_{26.9}$ SHNO, corresponding to the minimum $N^2$. These results confirm that the reduced $|\mu_0 M_{eff}|$ near the compensation point directly accounts for the narrow linewidth $\Delta f$ in the Pt/$Co_{73.1}Gd_{26.9}$ SHNO. Notably, since $M_{eff}$ is negative in this PMA system, Equation (3) shows that $N$ remains positive, indicating a propagating spin wave mode. This is further supported by micromagnetic simulations, in Section S7 of Supporting Information.

Because linewidth is closely related to phase noise, we further perform time-domain phase noise measurements using the zero-crossing technique (Section S8, Supporting Information). The measurements are performed on Co-rich ($Co_{83.1}Gd_{16.9}$), near-compensated ($Co_{73.1}Gd_{26.9}$), and Gd-rich ($Co_{57.0}Gd_{43.0}$) SHNOs, as illustrated in Figure 3e. The phase noise of the near-compensated device is nearly an order of magnitude lower than that of Co- or Gd-rich samples, consistent with their narrower $\Delta f$. Moreover, the current-frequency tunability ($df/dI_{DC}$), an indicator of nonlinearity for spin-oscillators [38], follows the same trend as $|\mu_0 M_{eff}|$ and also reaches its minimum near the compensation point (Section S9, Supporting Information). Meanwhile, time-domain measurements, together with fast Fourier transform and time-dependent frequency analyses, show a single temporally and spectrally stable auto-oscillation mode in the Pt/$Co_{73.1}Gd_{26.9}$ SHNO, excluding resolvable spectral mode competition as the origin of the linewidth narrowing (Section S8, Supporting Information).

We next evaluate the threshold current density ($J_{th}$) as a function of Gd concentration, as shown in Figure 4a. $J_{th}$ denotes the threshold current density at the onset of a detectable auto-oscillation signal, identified from the current-dependent power spectral density (PSD). Details of the $J_{th}$ extraction are provided in Section S5, Supporting Information. $J_{th}$ of Pt/$Co_{1-x}Gd_x$ SHNOs reaches a minimum near the compensation composition (Pt/$Co_{73.1}Gd_{26.9}$), indicating that near-compensated ferrimagnetic SHNOs achieve auto-oscillations at significantly lower current densities.

To investigate the origin of this reduction in $J_{th}$, we pattern the films into a bilayer microstrip and perform spin-torque ferromagnetic resonance (ST-FMR) measurements (Methods and Section S10, Supporting Information). The Gilbert damping constant ($\alpha$) is a key parameter characterizing magnetization dynamics. It is determined from the linear fit of the linewidth ($\mu_0\Delta H$) versus frequency [39-40], where each linewidth is obtained from Lorentzian fits to the ST-FMR spectra. The upper panel in Figure 4b summarizes the extracted $\alpha$ versus Gd concentration. According to the two-sublattice mean-field model, $\alpha$ is expected to diverge at the angular momentum compensation point ($x_{AMC}$), where the ratio of the magnetization to the gyromagnetic ratio for the two sublattices satisfies $M_{Co}/\gamma_{Co} = M_{Gd}/\gamma_{Gd}$ [41]. Therefore, a pronounced peak near $x = 24.5\%$ denotes $x_{AMC}$ [41]. Away from this point, $\alpha$ remains in the range 0.028 - 0.045 in the Pt (6 nm)/$Co_{1-x}Gd_x$ (7 nm) stacks, comparable to values in ferromagnetic SHNOs such as Pt/Py (0.024 - 0.028) [39, 42] and W/CoFeB/MgO (0.01 - 0.04) [15, 43]. Because $x = 26.9\%$ is closer to the reported $x_{MC}$ (25% - 26%) than to $x_{AMC}$ (~ 23%) [26, 44], the damping constant $\alpha$ remains moderate. This avoids the divergence of $\alpha$ near $x_{AMC}$, reduces energy dissipation from excessive damping, and enables a small $J_{th}$.

The damping-like torque efficiency ($\xi_{DL}$) and spin Hall conductivity ($\sigma_{SH}$) govern charge-to-spin conversion and determine the current needed for auto-oscillations. $\sigma_{SH}$ reflects the efficiency of charge-to-spin current generation, whereas $\xi_{DL}$ represents the spin-torque efficiency. Large $\sigma_{SH}$ and $\xi_{DL}$ enable efficient spin-torque excitation, resulting in a low $J_{th}$. $\xi_{DL}$ is obtained from DC bias-dependent ST-FMR measurements using [39, 45]

$$\xi_{DL} = \frac{2e}{\hbar}\frac{M_s w t_{CoGd}^2}{\sin\varphi}\mu_0\left(H_{res} + 0.5M_s\right)\frac{\Delta\alpha}{\Delta I_{DC}^{Pt}}, \tag{6}$$

where $w$, $t_{CoGd}$, $\mu_0$, $\varphi$, and $H_{res}$ denote the width of the ST-FMR device, the thickness of the $Co_{1-x}Gd_x$ layer, the vacuum permeability, the in-plane field angle, and the resonance field, respectively. The current in the Pt layer ($\Delta I_{DC}^{Pt}$) is evaluated from a parallel resistor model based on the measured resistivities of Pt and $Co_{1-x}Gd_x$. An example of $\alpha$ versus current in Pt under opposite magnetic-field polarities for the Pt/$Co_{73.1}Gd_{26.9}$ device is provided in Section S10, Supporting Information. As shown in the lower panel of Figure 4b, the DLT efficiency $\xi_{DL}$ (blue symbols) reaches a maximum of 0.102 for Pt/$Co_{73.1}Gd_{26.9}$, demonstrating enhanced SOT efficiency near the magnetization compensation point [25]. The corresponding spin Hall conductivity $\sigma_{SH}$ (red symbols) ranges from $2.1 \times 10^5$ to $3.7 \times 10^5$ $(\hbar/2e)$ $(\Omega\cdot m)^{-1}$, showing a similar maximum near $x_{MC}$. Meanwhile, the extracted field-like torque efficiency $\xi_{FL}$ is finite but smaller than the damping-like torque efficiency $\xi_{DL}$ (Section S10, Supporting Information), indicating that the anti-damping torque responsible for auto-oscillation is mainly governed by the damping-like SOT.

Using the experimentally determined $\alpha$, $M_s$, and $\sigma_{SH}$, we estimate $J_{th}$ from theoretical models [17, 38, 46-48]

$$J_{th} \propto \alpha M_s t_{CoGd} \left(M_s + H_{ext}\right) / \rho_{Pt} \sigma_{SH}, \quad (7)$$

where $\rho_{Pt}$ denotes the electrical resistivity of Pt. The calculated values (Figure 4a, red dashed line) reproduce the experimentally observed trend. As a result, the near-compensated Pt/$Co_{73.1}Gd_{26.9}$ SHNO, with high SOT efficiency, low $M_s$, and moderate $\alpha$, achieves $J_{th} = 1.01 \times 10^7$ A/cm$^2$ (Section S5, Supporting Information), which is an order of magnitude lower than the Pt/Py SHNO ($J_{th} = 1.64 \times 10^8$ A/cm$^2$, Section S5, Supporting Information). Notably, the low operating current not only reduces power consumption but also results in negligible Joule heating. The device temperature increases by only about 6 °C above room temperature at 0.1-0.25 mA and remains stable throughout operation (Section S11, Supporting Information).

Achieving auto-oscillation with a low or even zero external field is vital for developing compact on-chip hardware. In the Pt/$Co_{73.1}Gd_{26.9}$ SHNO, the low external field required for auto-oscillation is attributed to weak magnetic anisotropy, where the PMA field nearly cancels the out-of-plane demagnetizing field. From VSM measurements, the saturation magnetization $M_s$ is determined to be 36.01 emu/cm$^3$ (Figure 1c), corresponding to a demagnetizing field of approximately 45.3 mT, assuming a demagnetization factor $N_z \approx 1$. The PMA field is measured to be $\mu_0 H_k$ = 51.8 mT (Section S2, Supporting Information), only 6.5 mT greater than the demagnetizing field, indicating a near-easy-plane condition where internal anisotropies almost cancel along the $z$-axis [22].

In conventional ferromagnetic SHNOs, the internal magnetic field arises from the combined contributions of the anisotropy, demagnetizing, and external fields. Consequently, the anisotropy and demagnetizing fields can partially offset the external field. In contrast, in near-compensated SHNOs, the PMA and demagnetizing fields nearly cancel each other, making the external field more effective and thereby reducing the external field required to set the precessional axis of auto-oscillations (Section S12, Supporting Information). This operating principle is analogous to the observation in easy-plane SHNOs at cryogenic temperatures, where the applied field can be minimal because anisotropy and demagnetizing fields compensate each other [22]. Micromagnetic simulations (Methods) further support this behavior, revealing large precession cone angles of 6.71° to 10.25° for polar angles between 60° and 120° (Figure 5a), two orders of magnitude larger than those of conventional Pt/Py SHNOs (approximately 0.045°, Figure 5b), consistent with near-easy-plane precession under weak anisotropy.

Additional evidence supporting the near-cancellation of anisotropy in the near-compensated Pt/$Co_{73.1}Gd_{26.9}$ SHNO is shown in Figure 5c. We measure the $\theta$-dependent auto-oscillation frequency at fixed field magnitude, $\varphi$, and $I_{DC}$. The nearly constant frequency can be attributed to the effective cancellation of magnetic anisotropy, causing the external field to be equivalent to the internal field under magnetostatic boundary conditions (Section S12, Supporting Information). As a result, the auto-oscillation frequency, dependent on the magnitude of the internal field, is primarily governed by the magnitude of the external field. In Figure 5c, where the magnitude of the external field is constant, the auto-oscillation frequency remains nearly invariant in the Pt/$Co_{73.1}Gd_{26.9}$ SHNO, consistent with the analytical calculation (red line) and simulation (blue line). For comparison, the Co-rich Pt/$Co_{83.1}Gd_{16.9}$ SHNO ($\mu_0 M_{eff}$ = –215 mT) exhibits a sinusoidal $\theta$-dependence in both the experimental data and analytical fitting (Figure 5d) because both the anisotropy and external field contribute to the internal field (Section S12, Supporting Information). These observations indicate that zero $M_{eff}$ facilitates the realization of field-free spin oscillators.

The near-compensated $Pt/Co_{73.1}Gd_{26.9}$ SHNOs demonstrate a unique combination of low threshold current, minimal external field, and narrow linewidth. Near the magnetization compensation point, the key SHNO parameters, including the auto-oscillation linewidth $\Delta f$ and threshold current density $J_{th}$, show modest variation with Gd concentration, while the best performance is still obtained for the composition closest to the magnetization compensation point. As summarized in Figure 5e, the single-constriction $Pt/Co_{73.1}Gd_{26.9}$ SHNO exhibits auto-oscillations with a threshold current density $J_{th} = 1.01 \times 10^7$ A/cm$^2$ under an applied field of $\mu_0 H_{ext} = 5$ mT. These values correspond to a 16-fold reduction in $J_{th}$ and a tenfold reduction in the applied field compared with the Pt/Py control ($J_{th} = 1.64 \times 10^8$ A/cm$^2$, $\mu_0 H_{ext} = 50$ mT), and outperform other single-constriction ferromagnetic or hybrid SHNOs operating at room temperature. In addition to the reduced power consumption, the $Pt/Co_{73.1}Gd_{26.9}$ SHNO achieves a record-low linewidth of 0.61 MHz at 5 mT, standing out from other single-constriction SHNOs (Figure 5f). The results further suggest that materials with low saturation magnetization $M_s$, while simultaneously possessing enough perpendicular anisotropy to nearly cancel the demagnetization field (i.e., $M_{eff} \approx 0$), can achieve auto-oscillation with a very small applied current and field. Achieving such a balance is challenging in many room-temperature ferromagnets.

Compared with existing technologies [49-51], the near-compensated $Pt/Co_{73.1}Gd_{26.9}$ SHNOs demonstrate GHz operation with a sub-mA threshold current and µW-level input power, matching or even outperforming representative GHz CMOS- and SAW-based oscillators (Section S13, Supporting Information). These results highlight the potential of near-compensated ferrimagnetic SHNOs as energy-efficient nanoscale microwave sources, with future improvements in linewidth and field-free operation expected to further advance CMOS-compatible integration.

In thin-film SHNOs, PMA is needed to offset the demagnetizing field along the *z*-axis and thereby reduce the effective anisotropy. However, PMA systems often also possess a relatively large $M_s$, which further enhances the *z*-axis demagnetization term and therefore increases the anisotropy that the perpendicular component needs to counteract [22, 24]. Therefore, the near-compensated ferrimagnets provide a promising material platform for simultaneously reducing the required current and applied field, as they can combine a small $M_s$ and a weak effective anisotropy. These findings highlight the performance advantages of compensated ferrimagnets in both energy efficiency and signal quality, marking a substantial step forward in SHNO technology.

**3. Conclusion**

We have demonstrated that $Co_{1-x}Gd_x$ ferrimagnetic alloys provide an energy-efficient material platform for SHNOs. By tuning the composition, we optimize the sublattice-specific magnetic characteristics, identifying near-compensated Pt/$Co_{73.1}Gd_{26.9}$ as a high-performance candidate for SHNOs. The Pt/$Co_{73.1}Gd_{26.9}$ SHNOs achieve their minimum linewidth of 0.61 MHz at an operating current density of $1.43 \times 10^7$ A/cm$^2$ under a magnetic field of 5 mT. These results surpass conventional ferromagnetic SHNOs and highlight the intrinsic advantages of ferrimagnets, including the reduced effective magnetization, enhanced SOT efficiency, low nonlinearity, and suppressed phase noise. Moreover, we attribute the reduced external field to the weak anisotropy in near-compensated ferrimagnets, enabling low-field and potentially field-free operation, which is essential for practical on-chip integration.

Looking ahead, integrating near-compensated ferrimagnets with improved spin-source layers, optimized geometries, and device-scaling strategies provides a strong pathway toward further advances in energy-efficient, high-performance SHNOs. From a broader materials perspective, near-compensated ferrimagnets bridge ferromagnetic and antiferromagnetic spintronic oscillators by offering antiferromagnet-like spin dynamics while retaining practical electrical tunability and detection, highlighting a materials-design route that may extend SHNOs beyond conventional ferromagnets to ferri-, antiferro-, and potentially altermagnetic systems. These advancements will position spintronic oscillators for impactful deployment in next-generation wireless communications and unconventional computing, enabling new classes of energy-efficient, scalable electronic systems.

## 4. Experimental Section

*Sample Fabrication:* To prepare the $Co_{1-x}Gd_x$ films, a series of film stacks consisting of Ta (1 nm)/Pt (6 nm)/$Co_{1-x}Gd_x$ (7 nm)/$TaO_x$ (2 nm) were deposited on highly resistive (10 kΩ·cm) silicon substrates. The Co and Gd concentrations were adjusted during sputtering by varying the Gd sputtering power from 40 to 120 W. The deposition was performed at a base pressure below $3 \times 10^{-9}$ Torr and at room temperature, with the argon pressure maintained at 5 mTorr during the $Co_{1-x}Gd_x$ layer deposition. Devices were patterned into nano-constriction shapes using ma-N 2401 resist and an electron beam lithography system (Raith EBPG 5200) operating at 100 kV, followed by Ar-ion milling. Subsequently, electrodes composed of Ta (5 nm)/Cu (800 nm)/Ta (5 nm) were fabricated using a lift-off process. Patterning was achieved using an ultraviolet maskless lithography machine (TuoTuo Technology, UV Litho-ACA), and the metal layers of the electrodes were deposited by sputtering.

*ST-FMR Measurements:* ST-FMR measurements were performed at room temperature using a three-dimensional electromagnet probe station (MPS-C-350), which allowed precise external magnetic field control. A microwave signal from a signal generator (Agilent E8257D) was injected into the device through a bias tee with a power of 10 dBm. The generated DC voltage was measured with a lock-in amplifier (Stanford Research SR830). For DC-biased ST-FMR measurements, a Keithley 2400 source was used to apply the DC bias. The DLT efficiencies, extracted from the frequency dependence of the linewidth (i.e., from the slope of $\mu_0\Delta H$ versus $f$), represent frequency-averaged values.

*PSD and Time-domain Measurements:* All PSD measurements of SHNO auto-oscillations were conducted at room temperature using a three-dimensional electromagnet probe station (MPS-C-350) to apply an external magnetic field. A direct current was supplied to the SHNO through the DC port of a bias tee using a DC source (Keithley 2400). The microwave signals from the SHNO were amplified by a low-noise amplifier (LNA) with a gain of 36 dB and recorded using a spectrum analyzer (Agilent E4448A). For the time-domain measurements, we used a 20 GHz high-bandwidth oscilloscope (Tektronix DPO72004C) to measure 10 μs - 1 ms traces at 10 - 100 GS/s. The time-domain voltage signal was used to calculate the phase noise.

*Micromagnetic Simulations:* The micromagnetic simulations were conducted using mumax$^3$. First, the electric current distribution of an SHNO at the actual scale was calculated using resistivity values obtained from experiments. The current distribution was then imported into mumax$^3$. In the simulation model, a region of 4 μm × 4 μm × 7 nm was analyzed with a mesh grid of 1024 × 1024 × 1. The magnetic parameters, including the saturation magnetization (36.01 emu/cm$^3$ for $Co_{73.1}Gd_{26.9}$), Gilbert damping constant (0.042 for $Co_{73.1}Gd_{26.9}$), and spin Hall conductivity ($3.7 \times 10^5$ $(\hbar/2e)$ $(\Omega\cdot m)^{-1}$ for Pt/$Co_{73.1}Gd_{26.9}$), were set to the experimentally measured values. Magnetic anisotropy was aligned along the *z*-axis to reflect the PMA behavior. The time-domain output of the averaged magnetization vector was transformed into the frequency domain using a fast Fourier transform. To minimize the impact of the initial magnetization fluctuation on the analysis, only the data from 80 to 100 ns were extracted, with a step size of 0.01 ns.

## Figures

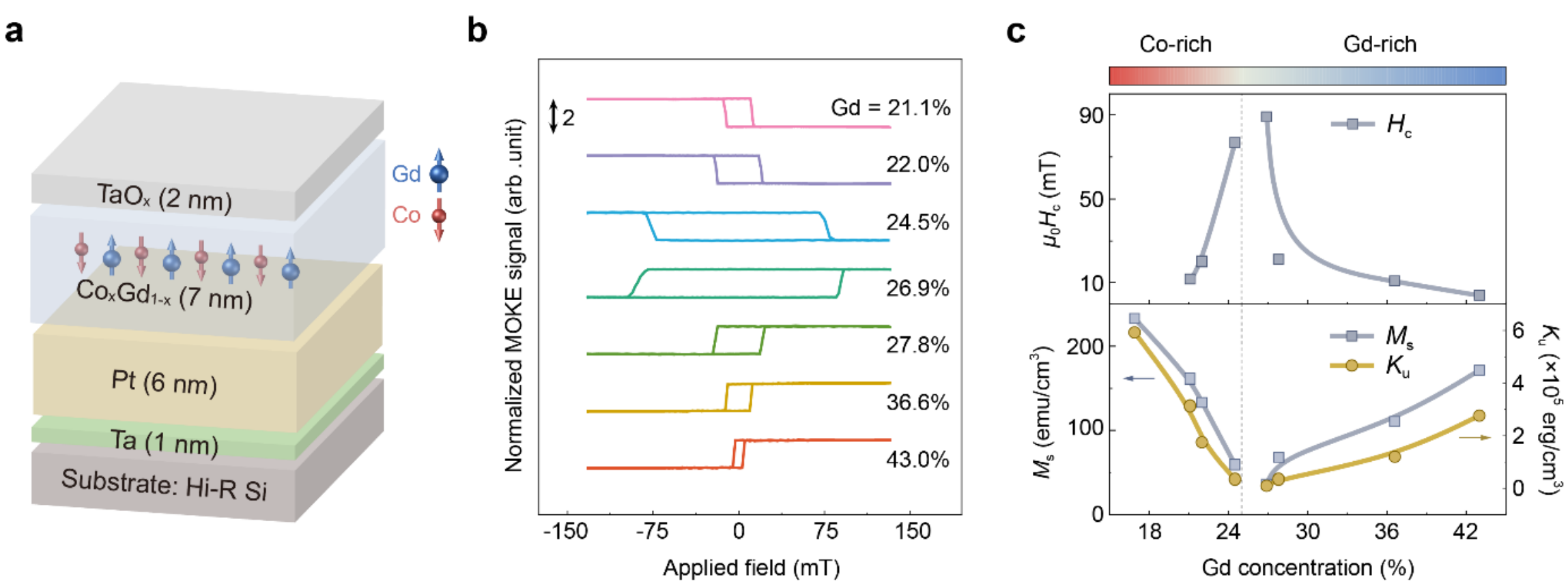

**Figure 1.** Composition-dependent magnetic properties of $Co_{1-x}Gd_x$ ferrimagnets. a) Schematic illustration of the film structure. The Co (red) and Gd (blue) sublattice moments within the $Co_{1-x}Gd_x$ layer are antiferromagnetically coupled. b) Out-of-plane static MOKE hysteresis loops measured at room temperature for the Pt (6 nm)/$Co_{1-x}Gd_x$ (7 nm) films with different Gd concentrations, showing a reversal of loop polarity between Gd concentrations of 24.5% and 26.9%, indicative of the transition from Co- to Gd-dominated magnetization. c) Composition-dependent coercive field ($H_c$), saturation magnetization ($M_s$), and uniaxial anisotropy energy density ($K_u$) for the Pt (6 nm)/$Co_{1-x}Gd_x$ (7 nm) films. The lines are drawn as a guide to the eye. Both $M_s$ and $K_u$ reach minima near $x \approx 25\%$ - 26% (the grey dashed line), indicating the magnetization compensation point. The color gradient above the panel denotes the transition from Co-rich to Gd-rich compositions, reflecting the changing relative abundance of Co and Gd atoms.

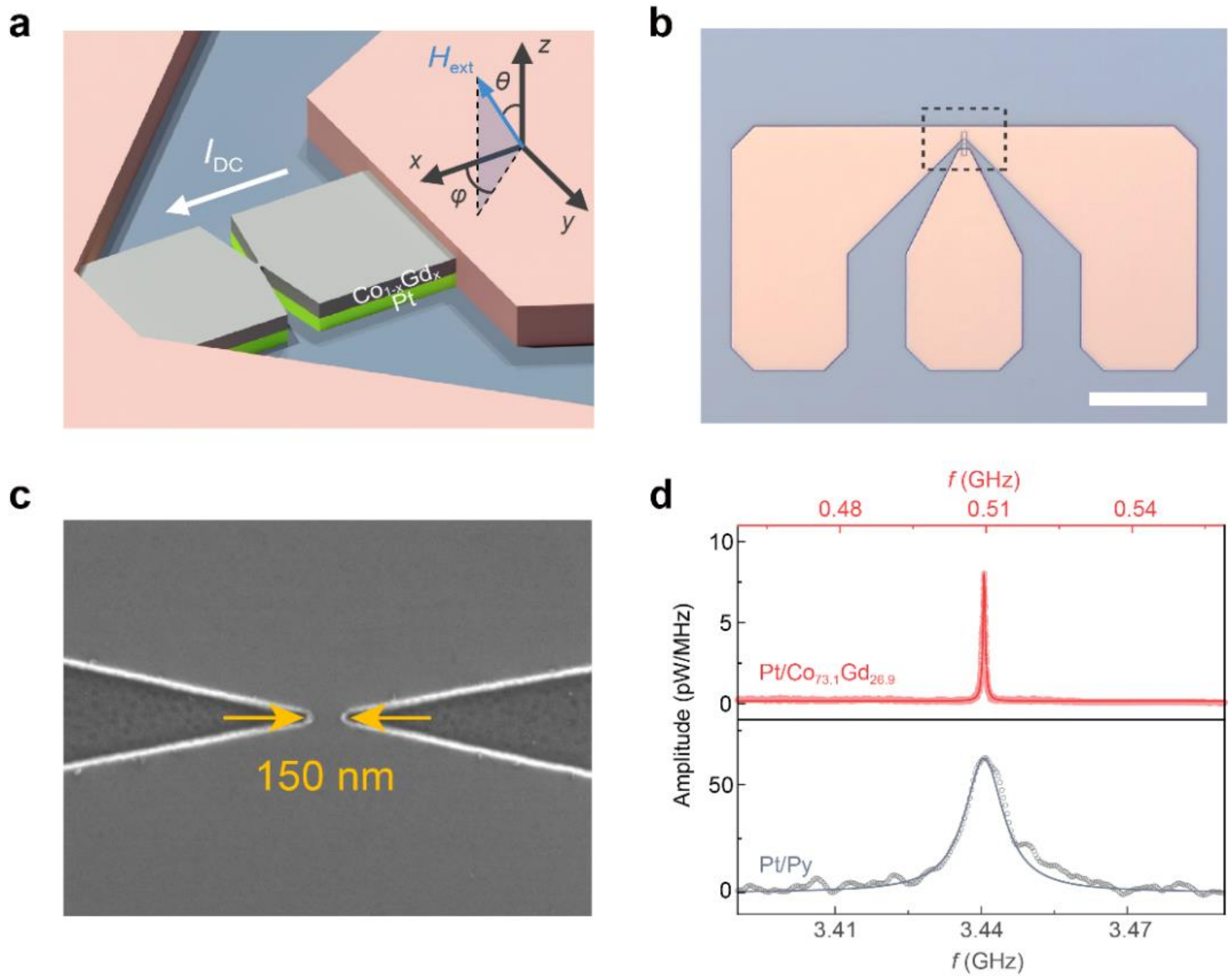

**Figure 2.** Device geometry and representative auto-oscillation spectra. a) Schematic of a nano-constriction SHNO consisting of a Pt (6 nm)/$Co_{1-x}Gd_x$ (7 nm) bilayer, with the direct current ($I_{DC}$) applied along the $x$-axis and the external magnetic field ($H_{ext}$) oriented by polar ($\theta$) and azimuthal ($\varphi$) angles. b) Optical micrograph of the fabricated SHNO integrated with a ground-signal-ground coplanar waveguide (scale bar, 100 μm). The SHNO is contained within the black dashed box. c) SEM top-view image of an SHNO with a 150 nm constriction. d) Auto-oscillation spectra of a near-compensated Pt/$Co_{73.1}Gd_{26.9}$ SHNO (upper panel), measured at $I_{DC}$ = 0.17 mA and $\mu_0 H_{ext}$ = 5 mT, and of a Pt/Py SHNO (lower panel), measured at $I_{DC}$ = 2.7 mA and $\mu_0 H_{ext}$ = 50 mT. Open circles show the measured spectra, and solid lines are Lorentzian fits.

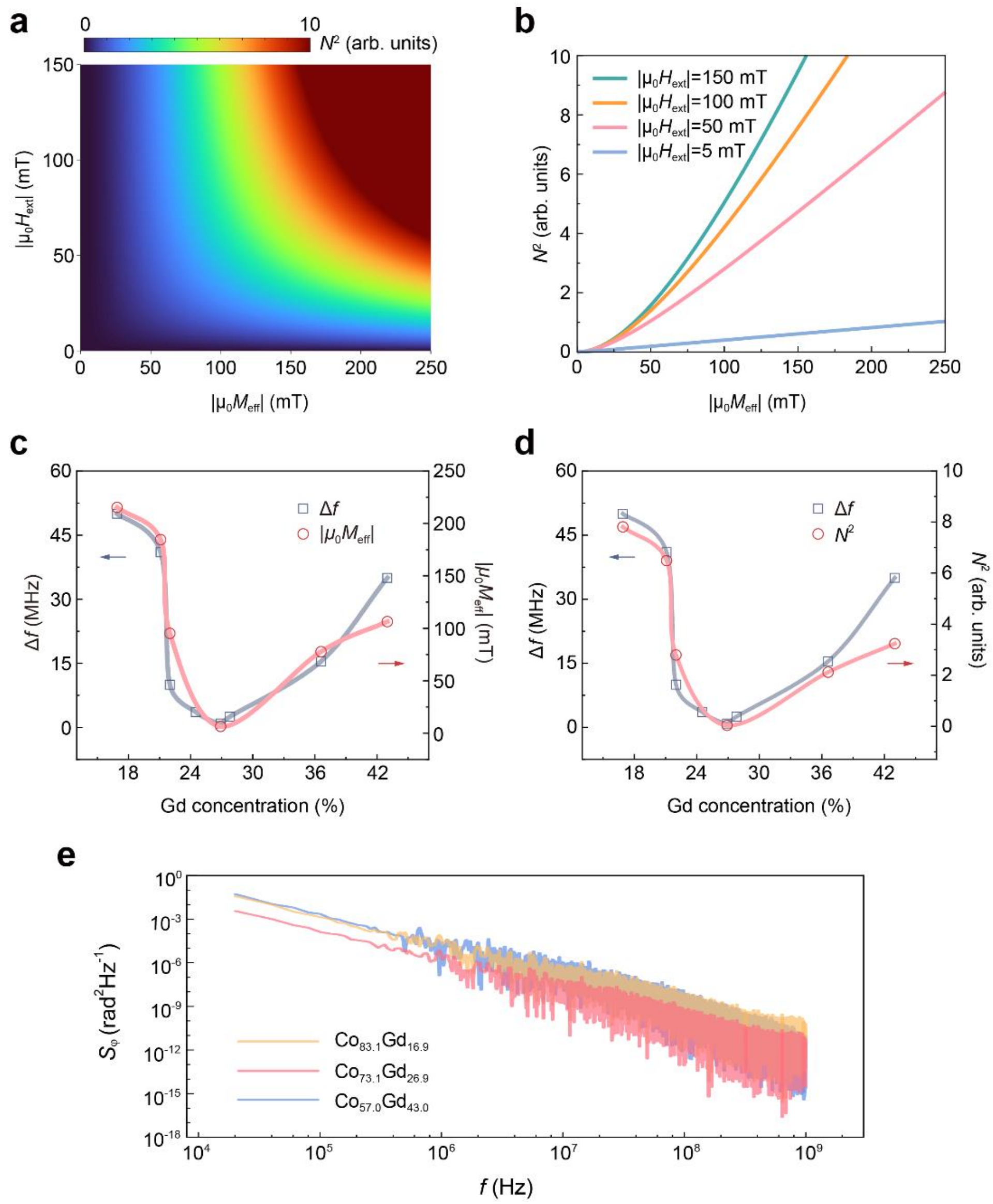

**Figure 3.** Dependence of $N^2$, linewidth, and phase noise on effective magnetization, external field, and Gd concentration. a) Color map of the squared nonlinear coefficient $N^2$ as a function of the effective magnetization magnitude $|\mu_0 M_{eff}|$ and the in-plane external magnetic field magnitude $|\mu_0 H_{ext}|$. b) Line cuts of $N^2$ extracted from panel (a) at fixed $|\mu_0 H_{ext}|$ values of 5, 50, 100, and 150 mT. c) Gd-concentration dependence of the linewidth $\Delta f$ and $|\mu_0 M_{eff}|$. d) Gd-concentration dependence of $\Delta f$ and $N^2$. The solid lines in panels (c,d) are guides to the eye. e) Phase noise spectrum ($S_\varphi$) for Co-rich ($Co_{83.1}Gd_{16.9}$), near-compensated ($Co_{73.1}Gd_{26.9}$), and Gd-rich ($Co_{57.0}Gd_{43.0}$) compositions, extracted from the time-domain measurements.

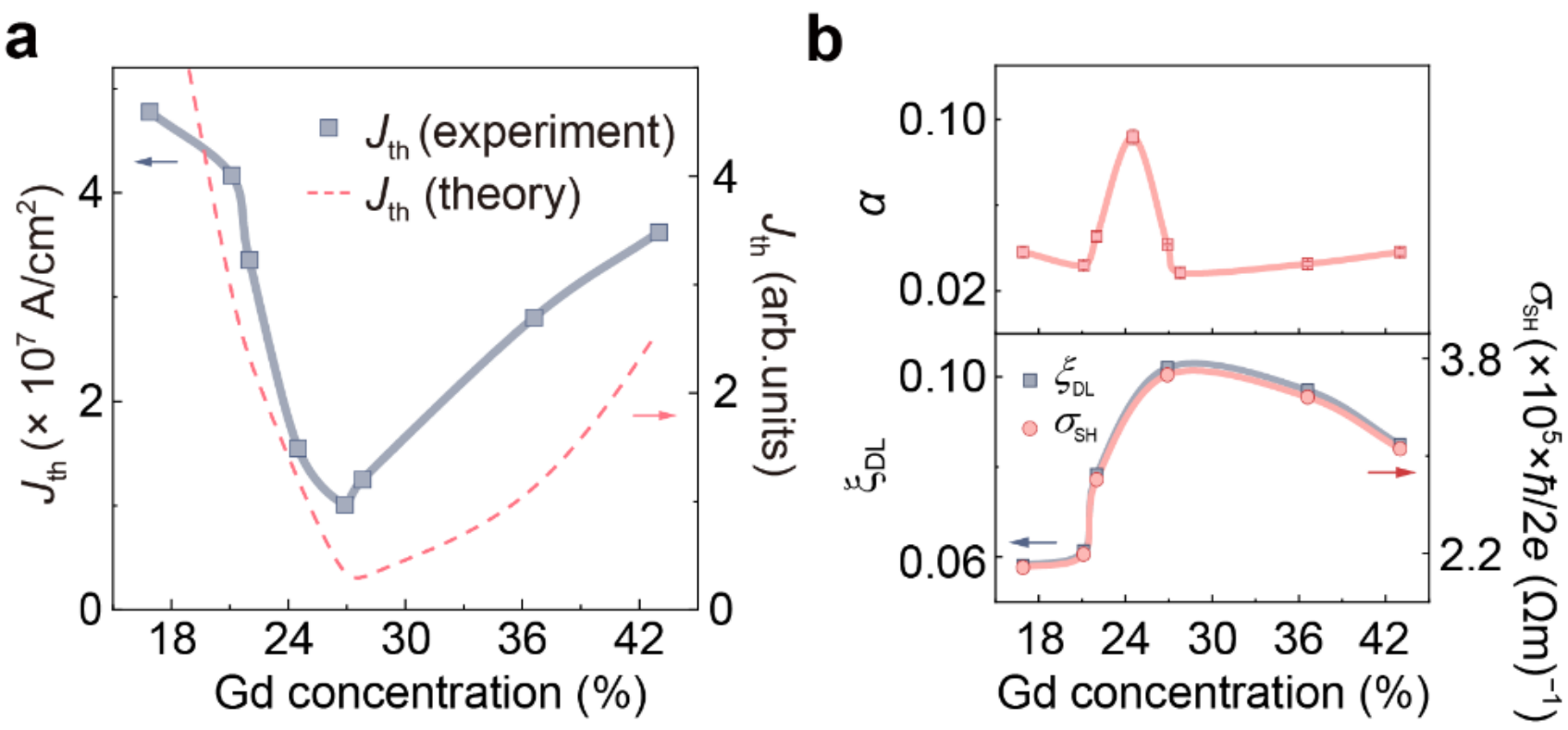

**Figure 4.** Influence of Gilbert damping and SOT efficiency on the threshold current density. a) Experimental and calculated auto-oscillation threshold current density ($J_{th}$) as a function of Gd concentration in Pt (6 nm)/$Co_{1-x}Gd_x$ (7 nm) SHNOs, with an external magnetic field $\mu_0 H_{ext}$ = 50 mT. b) Composition dependence of the Gilbert damping constant ($\alpha$, upper panel), damping-like torque efficiency ($\xi_{DL}$, lower panel), and spin Hall conductivity ($\sigma_{SH}$, lower panel). The lines are drawn as a guide to the eye.

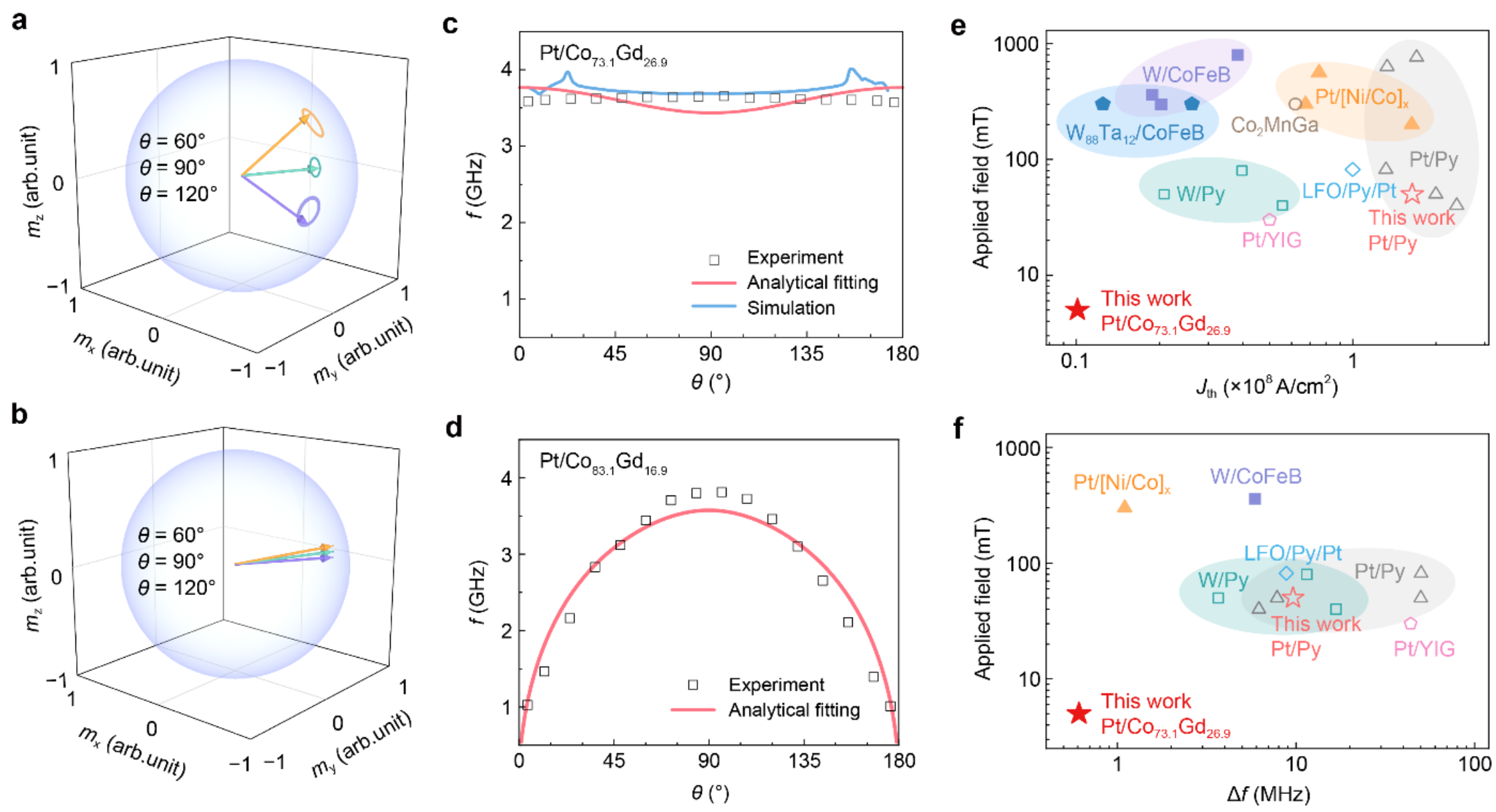

**Figure 5.** Field-dependent auto-oscillation in Pt (6 nm)/$Co_{1-x}Gd_x$ (7 nm) SHNOs and comparison with other ferromagnetic and hybrid SHNOs. a) Simulated magnetization trajectories of auto-oscillations in a Pt/$Co_{73.1}Gd_{26.9}$ SHNO at $\varphi = 66°$, for $\theta = 60°$ (yellow), $\theta = 90°$ (green), and $\theta = 120°$ (purple). b) Simulated magnetization trajectories of auto-oscillations in a Pt/Py SHNO under the same field conditions. Arrows indicate the normalized magnetization vectors in each case. c,d) Auto-oscillation frequency as a function of the field angle $\theta$ at fixed $\varphi = 66°$ for a Pt (6 nm)/$Co_{73.1}Gd_{26.9}$ (7 nm) SHNO (c) and a Pt (6 nm)/$Co_{83.1}Gd_{16.9}$ (7 nm) SHNO (d). Black symbols: experiment. Red lines: analytical fitting. Blue lines: micromagnetic simulation. The angular definitions of $\theta$ and $\varphi$ are given in Figure 2a. e) The auto-oscillation threshold current density ($J_{th}$) and the corresponding applied fields of single-constriction ferromagnetic and hybrid SHNOs [7, 12-13, 15, 17-18, 21, 23-24, 52-56] in comparison with our work. f) The minimum auto-oscillation linewidth ($\Delta f$) and the corresponding applied fields of single-constriction ferromagnetic and hybrid SHNOs [7, 17-18, 23-24, 42, 53-54] in comparison with our work. In both (e) and (f), filled symbols represent SHNOs with PMA layers, while open symbols represent SHNOs with in-plane magnetic anisotropy layers. The star symbols indicate data obtained from this work.